\documentclass[12pt]{article}
\usepackage{amsfonts}
\usepackage{amsmath}
\usepackage{amssymb}
\usepackage{graphicx}
\usepackage{color}
\usepackage[all, knot]{xy}
\usepackage{tikz}

\usepackage{ulem}

\usepackage[utf8]{inputenc}
\usepackage{epstopdf}
\usepackage[footnotesize]{caption}
\usepackage{amsthm}
\usepackage{enumitem}
\usepackage{mathrsfs}

\usepackage[margin=3cm]{geometry}

\def \be {\begin{equation}}
\def \ee {\end{equation}}
\def \bea {\begin{eqnarray}}
\def \eea {\end{eqnarray}}
\def \nn {\nonumber}

\def \rr {\raise.35ex\hbox{\small $\prime$}\kern-.17em{\mbox{\large $\imath$}}}

\def \dels {\partial\kern-.6em /\kern.1em}
\def \As {{A\kern-.5em / \kern.5em}}
\def \Ds {D\kern-.7em / \kern.5em}
\def \a {\alpha}

\def \ks {k\kern-.5em /}
\def \ls {l\kern-.5em /}

\def \s {\sigma}

\newcommand{\ci}[1]{}

\newcommand{\ba}{\begin{eqnarray}}
\newcommand{\ea}{\end{eqnarray}}
\newcommand{\bal}{\begin{align}}
\newcommand{\eal}{\end{align}}
\newcommand{\bay}[1]{\left(\begin{array}{#1}}
\newcommand{\eay}{\end{array}\right)}

\def\DJ{{\fontencoding{T1}\selectfont\char208}}

\newcommand{\hide}[1]{}

\newlist{axioms}{enumerate}{2}
\setlist[axioms,1]{label=\textbf{A\arabic{axiomsi}.}, ref=A\arabic{axiomsi}}
\setlist[axioms,2]{label=\textbf{A\arabic{axiomsi}\rlap{\myEnumCounter{axiomsii}}.},%
                   ref=A\arabic{axiomsi}\myEnumCounter{axiomsii},%
                   align=parleft,%
                   leftmargin=0em,%
                   itemsep=1.4ex,%
                   before={\stepcounter{axiomsi}}}

  \usetikzlibrary{decorations.markings}

\begin{document}

\begin{titlepage}
\begin{center}

\textbf{\LARGE
Spectral Form Factor\\ as an OTOC Averaged\\
 over the Heisenberg Group
\vskip.3cm
}
\vskip .5in
{\large
Robert de Mello Koch$^{a, b}$ \footnote{e-mail address: robert@neo.phys.wits.ac.za}, Jia-Hui Huang$^a$ \footnote{e-mail address: huangjh@m.scnu.edu.cn}, Chen-Te Ma$^{a,c}$ \footnote{e-mail address: yefgst@gmail.com}, and Hendrik J.R. Van Zyl$^b$ \footnote{e-mail address: hjrvanzyl@gmail.com}
\\
\vskip 1mm
}
{\sl
$^a$
Institute of Quantum Matter,\\
 School of Physics and Telecommunication Engineering,\\
 South China Normal University, Guangzhou 510006, Guangdong, China.
\\
$^b$
School of Physics and Mandelstam Institute for Theoretical Physics,\\
University of Witwatersrand, Wits, 2050, South Africa.
\\
$^c$
The Laboratory for Quantum Gravity and Strings,\\
 Department of Mathematics and Applied Mathematics,\\
University of Cape Town,Private Bag, Rondebosch 7700, South Africa.
}\\
\vskip 1mm
\vspace{40pt}
\end{center}

\begin{abstract}
We prove that in bosonic quantum mechanics the two-point spectral form factor can be obtained as an average of the two-point out-of-time ordered
correlation function, with the average taken over the Heisenberg group.
In quantum field theory, there is an analogous result with the average taken over the tensor product of many copies of the Heisenberg group, one copy for each field mode.
The resulting formula is expressed as a path integral over two fields, providing a promising approach to the computation of the spectral form factor.
We develop the formula that we have obtained using a coherent state description from the JC model and also in the context of the large-$N$ limit of CFT and Yang-Mills theory from the large-$N$ matrix quantum mechanics.
\end{abstract}
\end{titlepage}

\section{Introduction}
\label{sec:1}

Many-body quantum chaos \cite{Haake:2010} plays an important role in the strong coupling dynamics of systems in many fields of physics, including condensed matter physics, quantum information theory, and quantum gravity. Although one cannot rely on the characteristic exponential sensitivity to initial conditions when considering unitary quantum evolution (see for example \cite{Gharibyan:2019sag}), quantum chaotic systems share common features.
\\

\noindent
Two main criteria used to decide if a system exhibits quantum chaos or not, are the spectral form factor (SFF) \cite{Dyer:2016pou} and the out-of-time ordered correlation function (OTOC) \cite{Larkin:1969}. These probe the irregular dynamics and sensitivity to the initial conditions respectively.
The motivation for the SFF is rooted in random matrix theory \cite{Guhr:1997ve}. It is conjectured that a generic quantized system with a classical chaotic limit should exhibit the spectral statistics of a random matrix ensemble.  A concrete realization of this conjecture is that of Sinai's billiard \cite{Bohigas:1983er, Ho:2017nyc, Evers:2008zz}.
A more recent related proposal identifies the late time behavior of certain strongly coupled theories (including the Sachdev-Ye-Kitaev (SYK) model) \cite{Kitaev:2015,Polchinski:2016xgd} with a large number of degrees of freedom, captured in the two-point SFF
\begin{equation}
g_2(\beta, t)\equiv \frac{R_2(\beta, t)}{ R_2(0, t)} \ \ ; \ \ R_2(\beta, t) \equiv |\textnormal{Tr}\left(\exp(-\beta H-i Ht) \right)|^2
\end{equation}
with a dynamical form of random matrix universality \cite{Cotler:2016fpe}. In the above $\beta$ is the inverse temperature and $H$ is the Hamiltonian of the system.
\\

\noindent
In contrast to the SFF, the OTOC probes chaos at early time \cite{Larkin:1969,Almheiri:2013hfa}.
This should be due to the quantum uncertainty relation or losing infinitesimal perturbation in a local quantum system \cite{Gharibyan:2019sag}.
In chaotic systems, the OTOC exhibits exponential decay with rate $\lambda$ and converges to a persistent small value.
A semiclassical analysis shows that the rate $\lambda$ is naturally related to a Lyupanov exponent \cite{Larkin:1969}.
Under some natural assumptions, it is possible to prove the bound $\lambda\le 2\pi/\beta$ in the regularized OTOC \cite{Maldacena:2015waa}, revealing that $\lambda$ is an interesting quantity for theoretical considerations.
Note that it has been shown that the unregularized OTOC does not share the universal Lyapunov exponent with the regularized OTOC due to the sensitivity of the infrared regulator \cite{Stanford:2015owe,Chowdhury:2017jzb,Tsuji:2017fxs, Kamenev:1999zza,Chamon:1999zz,Kamenev:2009jj, Liao:2018uxa, Lee:1985zzc, Garcia-Garcia:2017bkg, Romero-Bermudez:2019vej}.
Further, protocols to measure the regularized OTOC have been given \cite{Swingle:2016var, Roberts:2016wdl, Yao:2016ayk}, applied to Jaynes-Cummings (JC) interactions \cite{Zhu:2016uws, Iwasawa:1995} and the Loschmidt echo \cite{Kurchan:2016nju, Zurek:2019}, and implemented
\cite{Garttner:2016mqj,Li:2017pbq} which confirms that $\lambda$ is an experimental observable. Taken together,
these facts establish the OTOC as a useful probe of quantum chaos.
\\

\noindent
Recently, the saturation of the OTOC at late times has been studied by connecting to spectral statistics through the correlation functions \cite{Gharibyan:2019sag}. This motivates the central question that we would like to address in this letter: {\it What is the relation between the spectral statistics and OTOC?} This question is key to relating the early and late time quantum behavior of a chaotic system.
An important result in this direction was obtained in \cite{Cotler:2017jue}, where it was argued that the SFF can be obtained as an average of the OTOC.
Since this is the key focus of our article, we will review the argument of \cite{Cotler:2017jue}.
\\

Consider a quantum system with dynamics in an $L$-dimensional Hilbert space.
Recall the average over $L\times L$ unitary matrices with the Haar measure is \cite{Weingarten:1977ya}
\bea
\int dA\ A^j_k A^{\dagger\, l}_m={1\over L}  \delta^j_m\delta^l_k\,.\label{NiceIdent}
\eea
The integral over $A$ is over all possible unitary operators on the Hilbert space.
In terms of the regularized two-point OTOC ($\rho\equiv\exp(-\beta H)$ is the density matrix)
\begin{equation}
O(t)\equiv \mathrm{Tr}\big( A(0)\sqrt{\rho}A^\dagger (t)\sqrt{\rho}\big)
\end{equation}
it is clear that
%
%
%
\bea
\int dA\ O(t)
&=&{1\over L}\int dA\ {\rm Tr}(A\sqrt{\rho}e^{-iHt}A^\dagger e^{iHt}\sqrt{\rho})
\nn\\
&=&{1\over L^2}{\rm Tr}(e^{iHt-\beta H/2}){\rm Tr}(e^{-iHt-\beta H/2})
\nn\\
&=&R_2(\beta/2,t),
\eea
where we used \eqref{NiceIdent} in the second equality.
This result achieves a direct link between spectral properties of a quantum system
(embodied in the SFF) and more intuitive notions of chaos like the Lyapunov exponent (visible in the OTOC).
\\

We generalize this result to bosonic quantum mechanics and quantum field theory.
This is non-trivial since the Hilbert space is infinite dimensional,
suggesting that we need to average over U($\infty$).
The corresponding Lie algebra is the set of all hermitian operators.
At classical level, the functions $W_n^s = x_1^{n+s-1}x_2^{s-1}$ with the Poisson bracket $\{ x_1,x_2 \}=1$
close the $W_\infty$ algebra \cite{Bakas:1990sh}.
This Lie algebra has an infinite number of generators and, hence, defines an infinite-dimensional group.
We need the quantum version obtained by replacing $x$ and $p$ with the canonical position operator $X$ and canonical momentum operator $P$.
Averaging over this enormous Lie group is complicated and obscures the link between the OTOC and SFF.
In the next section, we argue that this conclusion is too hasty: the relationship between the OTOC and SFF only requires an average over the Lie group generated by $\vec{X}$ and $\vec{P}$.
This dramatic simplification has conceptual implications: it is more feasible to find a simple interpretation for averaging over what is essentially classical phase space \cite{Zhuang:2019jyq}, than averaging over Hilbert space.

\section{Main Result}
\label{sec:2}
The Heisenberg group is a two-dimensional Lie group generated by $X$ and $P$, with the usual Lie algebra
$[P,X]=-i$.
A general element of the group is specified by the variables, $q_1,q_2$, as follows
\begin{equation}
U(q_1,q_2)\equiv \exp(iq_1 X+i q_2 P).
\end{equation}
We have $U(q_1,q_2)U^{\dagger}(q_1,q_2) =1$.
By direct computation, we find
%
%
\bea
&&\int_{-\infty}^{\infty} {dq_1\over 2\pi} \int_{-\infty}^{\infty} dq_2\ \langle x_1 |U(q_1,q_2)|x_2\rangle \langle y_1|U^{\dagger}(q_1,q_2)|y_2\rangle
\nn\\
&=&\delta (x_2-y_1)\,\delta (x_1-y_2)\,,\label{NicePos}
\eea
which is precisely the analog of \eqref{NiceIdent} \cite{Cotler:2017jue,Weingarten:1977ya}
What we obtained precisely follows from the properties:
\bea
\exp(iq X)|x\rangle = \exp(iq x)|x\rangle;\qquad \exp(iqP) |x\rangle = |x-q\rangle.
\eea
 It is simple to repeat the computation in any convenient basis. For applications in higher dimensions it is necessary to include multiple copies of the Heisenberg group element.
\\

The formula \eqref{NicePos} already implies that the SFF \cite{Dyer:2016pou} is obtained as an average of the two-point OTOC \cite{Cotler:2017jue}
\bea
O(x,t, q_1, q_2)
=\langle x |U(q_1,q_2)\exp(-iHt)U^{\dagger}(q_1,q_2) \exp(iHt)|x\rangle
\eea
 over the Heisenberg group
%
\bea
&& \int_{-\infty}^\infty dq_1\int_{-\infty}^\infty {dq_2\over 2\pi}\int_{-\infty}^\infty dx\, O(x, t, q_1, q_2)
\nn\\
&=&\int_{-\infty}^\infty dx \int_{-\infty}^\infty dx_1\
\langle x_1|e^{-iHt}|x_1\rangle
\langle x|e^{iHt}|x\rangle\,.
\eea
As an illustrative example, consider the harmonic oscillator. The Hamiltonian of the harmonic oscillator is
\bea
H_{\mathrm{HO}}=\frac{P^2}{2}+\frac{\omega^2X^2}{2},
\eea
 where $\omega$
 is the frequency.
Using the spectrum of a harmonic oscillator, it is simple to obtain
\bea
R_2(0, t)=\frac{1}{2-2\cos (\omega t)}.
\eea
The solution to the Heisenberg equation of motion is
\bea
P(t)&=&-\omega X(0)\sin(\omega t)+P(0)\cos(\omega t);
\nn\\
X(t)&=&X(0)\cos(\omega t)+P(0)\sin(\omega t)/\omega,
\eea
 where $X(0)$ and $P(0)$ are Schr\"odinger picture operators. Hence the two-point OTOC is given by
%
\bea
&&\langle x |U(q_1,q_2)e^{-i H_{\mathrm{HO}}t}U^{\dagger}(q_1,q_2) e^{iH_{\mathrm{HO}}t}|y\rangle
\nn\\
&=& e^{-i (x+q_2)\left(q_1 \cos (\omega t)-q_2\omega \sin (\omega t)-q_1\right)}
\delta\left( x+q_2-y-{q_1\over\omega}\sin (\omega t)-q_2\cos (\omega t)\right)
\nn\\
&&\times e^{i\frac{(q_1^2-\omega^2 q_2^2)}{4\omega}\sin (2t\omega)
+i{q_1q_2\over 2}\cos (2t\omega)-i{q_1 q_2\over 2}}\,.
\eea
Given this explicit OTOC, we can verify that the SFF is obtained as an average over the Heisenberg group.
The average is integrating over $q_1,q_2$ (with a $1/2\pi$ factor) and the trace is an integral over $x$.
The integral over $q_2$ is performed using the delta function.
The remaining two integrals are Gaussian integrals and easily performed.
Although we have only discussed the $\beta=0$ case for simplicity, we have verified the connection for $\beta\ne 0$.
\\

Rewrite this computation in terms of oscillators since this generalizes easily to non-interacting scalar field theory, which is an assembly of
non-interacting oscillators.
Using
\bea
a=\frac{(P-i\omega X)}{\sqrt{2\omega}};\qquad  a^\dagger =\frac{(P+i\omega X)}{\sqrt{2\omega}},
\eea
the unitary operators that we have considered are given by
\bea
U(q_1,q_2)=e^{a\left(iq_2\sqrt{\omega\over 2}-{q_1\over \sqrt{2\omega}}\right)}
e^{a^\dagger \left(iq_2\sqrt{\omega\over 2}+{q_1\over \sqrt{2\omega}}\right)}
e^{{q_1^2\over 4\omega}+{q_2^2\omega\over 4}}\nonumber\,.
\eea
The relation between the two-point OTOC and SFF is as before.
Now consider a non-interacting scalar field theory, in a box (with a periodic boundary condition), so that momenta $\vec{k}$ are discrete, with an oscillator for every $\vec{k}$. The Hamiltonian is
\bea
H_{\mathrm{NS}}=\frac{1}{V}\sum_{\vec{k}}\ \frac{1}{2} \tilde{a}^{\dagger}(\vec{k})\tilde{a}(\vec{k}),
\eea
where $V$ is the volume of the box. The $\tilde{a}^{\dagger}$ and $\tilde{a}$ are the usual creation and annihilation operators in the box, and they satisfy the commutation relation
\bea
\lbrack \tilde{a}(\vec{k}_1), \tilde{a}^{\dagger}(\vec{k}_2)\rbrack=2V\omega_{\vec{k}_1}\delta_{\vec{k}_1\vec{k}_2},
\eea
 where
\bea
\omega_{\vec{k}_1}^2\equiv |\vec{k}_1|^2+m^2
\eea
 with $m$ the mass of the non-interacting scalar field. Hence we can perform the field redefinition
\bea
\tilde{a}(\vec{k})\equiv \sqrt{2V\omega(\vec{k})}a(\vec{k})
\eea
 and apply the result of the harmonic oscillator to the non-interacting scalar field theory.
\\

The relevant unitary operator is
%
%
\bea
&&U\lbrack q_1(\cdot),q_2(\cdot)\rbrack
\nn\\
&=&
e^{\sum_{\vec k}a_{\vec k}\left(iq_2(\vec k)\sqrt{\omega_{\vec k}\over 2}
-{q_1(\vec k)\over \sqrt{2\omega_{\vec k}}}\right)}
e^{\sum_{\vec p}a^\dagger_{\vec p} \left(iq_2(\vec p)\sqrt{\omega_{\vec p}\over 2}
+{q_1(\vec p)\over \sqrt{2\omega_{\vec p}}}\right)}
e^{\sum_{\vec l}\left( {q_1^2(\vec l)\over 4\omega}+{q_2^2(\vec l)\omega\over 4}\right)}\,.
\nn\\
\eea
The unnormalized two-point SFF written at $\beta=0$ for simplicity, for the non-interacting scalar field, is given by
%
%
\bea
&&\int_{-\infty}^{\infty} \prod_{\vec k} {dq_1(\vec k)\, dq_2(\vec k)\over 2\pi}\
 {\rm Tr}
\left( U\lbrack q_1,q_2\rbrack e^{-iH_{\mathrm{NS}}t}U^{\dagger}\lbrack q_1,q_2\rbrack e^{iH_{\mathrm{NS}}t}\right)
\nn\\
&=&\prod_{\vec k}{1\over \left( 2\sin {t\omega_{\vec k}\over 2}\right)^2}\,.
\eea
We are now using a square bracket notation to stress that $U$ is a functional of $q_1(\vec k)$
and $q_2(\vec k)$, that is, that $U[q_1(\cdot),q_2(\cdot)]$ depends on the functions $q_1(\vec k),q_2(\vec k)$ and not just
their values at some momentum $\vec k$.
Notice that we are integrating over all possible values of each of the variables $q_1(\vec k)$ and $q_2 (\vec k)$, for
every possible $\vec k$, that is, we are performing a path integral over the fields $q_1(\cdot)$ and $q_2(\cdot)$.
We could write the averaged OTOC as
\bea
\int [Dq_1]\int [Dq_2]\ {\rm Tr}
\big( U\lbrack q_1,q_2\rbrack e^{-iH_{\mathrm{NS}}t}
U^{\dagger}\lbrack q_1,q_2\rbrack e^{i H_{\mathrm{NS}}t}\big)\,.
\eea
This formula for the SFF in quantum field theory is a promising starting point: any of the standard approximation techniques used
to study the path integral can be applied.

\section{Late-Time Limit: Coherent State and Large-$N$}
\label{sec:3}
For applications to black hole physics, we want the late time behavior of the SFF which is a probe
of information loss \cite{Dyer:2016pou, Maldacena:2001kr}.
Typically a large time limit is a classical limit of sorts \cite{Berry:1979in}, conveniently described with coherent states (with minimum uncertainty) \cite{Hepp:1974vg}.
Our first example is a model from quantum optics, the JC model, and we develop
the coherent state language to study the averaged OTOC.
Our choice of model is guided by the fact that protocols to measure the OTOC in the JC model have been advanced
in \cite{Zhu:2016uws}.
Another setting in which it would be desirable to have a better understanding of the SFF is large-$N$ CFT and
large-$N$ Yang-Mills theory, which is interesting in the quantum gravity community.
For these theories, the planar limit is a theory of generalized free fields, and our simple formulas developed
for the oscillator are applicable.
We demonstrate this with a simple example from matrix quantum mechanics: the planar limit of the oscillator
perturbed by a quartic potential.
\\

We consider the exactly solvable model from the two-photon non-degenerate JC model with the rotating wave approximation, which ignores the oscillating fast term \cite{Iwasawa:1995}. The effective Hamiltonian is $$H_{\mathrm{JC}}\equiv N_1+N_2+M,$$ where
\bea
N_j &=& \omega_j \left(a_j^\dagger a_j + \frac{1}{2}(\sigma_z+1) \right);
\nn\\
M &=&  \frac{1}{2}\Delta (\sigma_z+1) + g_a (a_1 a_2 \sigma^+ + a_1^\dagger a_2^\dagger\sigma^-) .
\eea
The frequency of two photons $\omega_1$ and $\omega_2$ coincide perfectly with the cavity modes.
The $\sigma$'s are Pauli spin operators for the two-level atom of frequency $\omega_0$.
The parameter $g_a$ is the atom-field coupling constant.
The frequency difference between the two modes is
\bea
\delta = \omega_1-\omega_2,
\eea
 and the detuning parameter is
\bea
\Delta = \omega_0-(\omega_1+\omega_2).
\eea
Note that
\bea
[N_1,N_2]=0;\qquad [N_1,M]=0,;\qquad [N_2,M]=0.
\eea
 The exact spectrum of this model can be solved for \cite{Iwasawa:1995}, and it is thus straightforward to compute the SFF.
For this model we wish to show that the averaged OTOC again reproduces the SFF and showcase that the coherent state language is natural for the computation.  The coherent states that we use are:
\bea
 &&a_1|\alpha_1\alpha_2\rangle =\alpha_1|\alpha_1\alpha_2\rangle;\qquad
 a_2|\alpha_1\alpha_2\rangle =\alpha_2|\alpha_1\alpha_2\rangle;
\nn
\eea
\bea
 |\alpha_1\alpha_2\rangle
=\exp\big(-(|\alpha_1|^2+|\alpha_2|^2)/2\big)\exp\big(\alpha_1 a_1^\dagger+\alpha_2 a_2^\dagger\big)|0,0\rangle.
\eea
 Completeness of the coherent states is
\bea
\int {d^2\alpha_1\over \pi}\int {d^2\alpha_2\over \pi}\,\,
|\alpha_1\a_2\rangle\langle\alpha_1\alpha_2|=1\,.
\eea
In terms of the unitary operator
\bea
U(q_1,q_2,r_1,r_2)=\exp(iq_1 X_1+iq_2 P_1+ir_1 X_2+ir_2P_2),
\nn\\
\eea
we compute the regularized two-point OTOC (repeated indices $a,b$ are summed over 1,2)
\bea
\label{C(t)}
&&C(t)
\nn\\
&=&\langle \alpha_1\alpha_2 |U(q_1,q_2,r_1,r_2)[e^{-\beta H_{\mathrm{JC}}/2 -iH_{\mathrm{JC}}t}]_{aa}
\nn\\
&&\times U(q_1,q_2,r_1,r_2)^\dagger
 [e^{-\beta H_{\mathrm{JC}}/2+i H_{\mathrm{JC}}t}]_{bb}|\alpha_1\alpha_2\rangle\,,
\eea
where $[\cdots]_{aa}$ is the matrix element of the row-$a$ and the column-$a$ with the repeated summation. To evaluate the OTOC we insert as many of the coherent state identity resolutions as required.
Direct computation gives
%
%
\bea
\label{nice}
&&\langle \alpha_1\alpha_2 |U(q_1,q_2,r_1,r_2)|\gamma_1^1\gamma_2^1\rangle
\nn\\
&=&e^{\bar\alpha_1 \left(iq_2\sqrt{\omega_1\over 2}+{q_1\over \sqrt{2\omega_1}}\right)}
e^{\bar\alpha_2 \left(ir_2\sqrt{\omega_2\over 2}+{r_1\over \sqrt{2\omega_2}}\right)}
\nn\\
&&\times
e^{\gamma^1_1\left(iq_2\sqrt{\omega_1\over 2}-{q_1\over \sqrt{2\omega_1}}\right)
+\gamma^1_2\left(ir_2\sqrt{\omega_2\over 2}-{r_1\over \sqrt{2\omega_2}}\right)}
e^{-{|\alpha_1|^2+|\alpha_2|^2+|\gamma_1^1|^2+|\gamma_2^1|^2\over 2}+\bar\alpha_1\gamma_1^1+\bar\alpha_2\gamma_2^1}
\nn\\
&&\times
e^{-{q_1^2\over 4\omega_1}-{q_2^2\omega_1\over 4}-{r_1^2\over 4\omega_2}-{r_2^2\omega_2\over 4}}\,.
\eea
This matrix element is common for any two-particle problem - it is the coherent state expectation value of an element of the two-particle Heisenberg group.
The dependence on the coherent state parameters is noteworthy: the integrations that we need to perform over coherent state parameters are Gaussian integrals, which is a nice simplification that will be present in any problem.
We also suggest an interpretation for the averaging over the Heisenberg group: the average treats classical
phase space points as an ensemble for collecting statistical data about the dynamics.
This interpretation is used below to understand why the SFF develops a plateau at late times.
\\

By performing the prescribed average for the OTOC we recover exactly the SFF.  Exact computation is no doubt because the model is solvable.
In general, we will not be able to carry things out exactly.
Nevertheless, given that $t$ is a large parameter, the final integrations naturally lend themselves to saddle point evaluations.
\\

One of the settings in which we expect that our result will have immediate applications is  the large-$N$ limits of
conformal field and gauge theories.
As an illustrative example, consider large-$N$ matrix quantum mechanics.
In this case, the large-$N$ theory enjoys factorization \cite{Itzykson:1979fi} which can be exploited to simplify the analysis.
Concretely, consider the model
\bea
H_{\mathrm{QMN}}= \frac{P^j P^j}{2}+ \mu^2\frac{X^j X^j}{2}+g\frac{(X^j X^j)^2}{4},
\eea
where $j=1,2,\cdots,N$, and $g$ is the coupling constant.
Using the simplifications of the large-$N$, we replace this Hamiltonian with the approximate form ($\sigma$ is a constant.)
\bea
H_{\mathrm{QMNM}}=\frac{P^j P^j}{2}+\mu^2\frac{X^j X^j}{2}+\lambda\sigma \frac{X^j X^j}{2}.
\eea
The ’t Hooft coupling constant $\lambda \equiv gN$ is fixed as we scale $N\to\infty$, and we determine
\bea
\sigma = \frac{\sum_{j=1}^N\langle X^j X^j\rangle}{N}
\eea
from the two-point function. Repeated indices in the Hamiltonian are summed.
The large-$N$ theory is harmonic oscillators but now with a modified frequency.
To determine $\sigma$, note that the large-$N$ Schwinger-Dyson equation for the two-point function is
\bea
\left({d^2\over dt^2}+\mu^2 +\lambda\sigma\right) \sum_{j=1}^N\langle X^j(t)X^j(t')\rangle =-iN\delta (t-t')\,.
\eea
Therefore, we obtain the following:
%
%
\bea
&&\langle X^j(t)X^j(t')\rangle
\nn\\
&=&\int {d\omega\over 2\pi}{iN\over \omega^2-\mu^2-\lambda\s +i\epsilon}e^{i\omega (t-t')}
\nn\\
&=&{N\theta (t-t')\over 2\sqrt{\mu^2+\lambda\sigma}}e^{-i\sqrt{\mu^2+\lambda\sigma}\, (t-t')}
+{N\theta (t'-t)\over 2\sqrt{\mu^2+\lambda\sigma}}e^{i\sqrt{\mu^2+\lambda\sigma}\, (t-t')}\,.
\eea
Setting $t-t'=\epsilon$ and then choosing the vanishing limit $\epsilon\to 0^+$, we have
\bea
\sigma =\frac{1}{2\sqrt{\mu^2+\lambda\sigma}},
\eea
 which is easily solved.
In terms of this $\sigma$, the two-point SFF can be computed exactly
\bea
g_2(\beta, t)
=
\bigg(\frac{1+e^{-2\sqrt{\mu^2+\lambda\sigma}\beta}
-2e^{-\sqrt{\mu^2+\lambda\sigma}\beta}}
{1+e^{-2\sqrt{\mu^2+\lambda\sigma}\beta}
-2\cos (\sqrt{\mu^2+\lambda\sigma}t)e^{-\sqrt{\mu^2+\lambda\sigma}\beta}}\bigg)^N
\eea
for $\beta\ne 0$.
We have found good qualitative agreement between the large-$N$ result and numerical result for $N = 3$. The result showcases how the methods of large-$N$ can be used to simplify the analysis, and we expect to find applications for our result within this setting.

\section{Outlook}
\label{sec:4}
Our basic result is a sharp insight that touches on a long-standing open question:  the old conjecture that a generic classically chaotic system, when quantized, exhibits the spectral characteristics of the random matrix ensemble consistent
with its symmetries.
There are many possible extensions.
One could consider large dimension operators (with dimension scaling as $\mathcal{O}(N^2)$) in a large-$N$ CFT
with a gravity dual.
The computation of the SFF in this setting is related to the black hole information problem \cite{Papadodimas:2015xma}.
Another direction to explore is the long-time limit of the SFF.
On short time scales (small $t$) the factor $\exp(\pm i Ht)$ oscillates slowly on phase space.
The averaging over the Heisenberg group uses comparatively narrow Gaussians, and this does not remove many details of
the oscillations.
For large times, the oscillations of $\exp(\pm i Ht)$ are very rapid and the Gaussians are comparatively broad so
that they smooth away all the details that would be present in the $\exp(\pm iHt)$ oscillations, providing an explanation
of why the SFF approaches a constant value at late times - the so-called plateau \cite{Dyer:2016pou, Cotler:2016fpe}.
Our formula may provide insight into the time scale at which we transition to the plateau, something that deserves
further study.
Finally, if we interpret the average over the Heisenberg group as treating classical phase space points as an ensemble for collecting statistical data about the dynamics, it is natural to ask if, for an ergodic system, the OTOC \cite{Larkin:1969} might be self-averaging. In this case, one might be able to get the SFF from the OTOC of two ``typical'' unitary operators, rather than having to average.

\section*{Acknowledgments}
We would like to thank Izak Snyman and Beni Yoshida for penetrating insights that improved our understanding.
This work of RdMK and Van Zyl were supported by the South African Research Chairs
Initiative of the Department of Science and Technology and National Research Foundation
as well as funds received from the National Institute for Theoretical Physics (NITheP).
J.-H. Huang was supported by the Natural Science Foundation of Guangdong Province (No.2016A030313444).
CTM was supported by the Post-Doctoral International Exchange Program and would like to thank Nan-Peng Ma for his encouragement.
We would like to thank the Jinan University and Yukawa Institute for Theoretical Physics at the Kyoto University.
Discussions during the workshops, ``Jinan University Gravitational Frontier Seminar'' and ``Quantum Information and String Theory'', were useful to complete this work.


\end{document}